\documentstyle[12pt, epsf]{article}
\begin{document}

\begin{center}
\Large{\bf Time evolution of the two-jet events $e^+-e^- \to hadrons$
in the Dynamical String Model}\\[50pt]

\normalsize
  $^a$B. Iv\'anyi, $^a$Zs. Schram, $^a$K. Sailer
   and $^b$W. Greiner\\[25pt]

 $^a$Department for  Theoretical Physics,\\
   Kossuth Lajos University,\\
  H-4010 Debrecen, Pf. 5, Hungary\\[15pt]
 $^b$Institute for Theoretical Physics,\\
   Johann Wolfgang Goethe University,\\
  D-60054 Frankfurt am Main 11, Postfach 111932, Germany\\

\end{center}

\begin{abstract}

The time dependent pion emission rate and the mean life time
of the pion source in the center of mass frame of the two-jet
events have been determined for $e^+-e^- \to hadrons$ at
$\sqrt{s} = 20 - 50$ GeV by the Dynamical String Model.
It was established that the time needed for the creation
of the pion source is $5$ fm/c, whereas its life time
is $\tau_0 \approx 7 - 13$ fm/c for energies $\sqrt{s}
= 20 - 50$ GeV, respectively.

\end{abstract}

\section*{Introduction}

The purpose of this work is to obtain the reliable time evolution
of the hadronization process for two-jet events  $e^+ e^- \to hadrons$
at the energies $\sqrt{s} \approx 20-50$ GeV
in the framework of the Dynamical String Model.
This is achieved by making use of the particular sensitivity
of the simulated Bose-Einstein correlation to the decay constant
of the hadronic string. We show that it is possible to arrive at
an agreeable quantitative description of both the Bose-Einstein
correlation and the single particle data by an appropriate choice
of the string decay constant, being close to its value predicted
in Ref.\cite{nyolcb}. Having settled the time scale of the
hadronization process in this way, we determine the time
dependent emission rate and the mean life time of the
pion source.

The Dynamical String Model was developed in order
 to describe
high energy hadronic processes\cite{elso}. In the model the hadrons are
represented by classical open strings and their free motion is governed
by the Nambu-Goto action. Two types of string interactions are introduced.
The excited hadronic strings decay due to  the
quark-antiquark pair creation in the chromoelectric field of the
flux tube represented by
the string. The collision of hadronic strings is modelled
by the
arm-exchange mechanism called rearrangement. It is assumed that the
strings  move freely between two interactions.
In addition to, hadronic strings have a finite transverse size ensuring
the right order of magnitude of the hadronic cross sections.

The parameters of the model are as follows.
The string tension $\kappa \approx 1$ GeV$/$fm is determined
 from the slope of
the leading Regge trajectory of hadron resonances.
The  radius $R$ of hadronic strings
 is fitted to the
proton-proton total cross section. The third parameter of the model,
the decay constant of the hadronic strings  can be expressed in terms
of the string radius and the
string tension making use of the analogy of the hadronic string with
the chromoelectric flux tube
\cite{masodik}. Recently, there have been given
theoretical arguments in favour of this analogy \cite{harmadik}.
Nevertheless, such an analogy does not lead to an unambiguous
relationship between the string tension $\kappa$ and the chromoelectric field
strength $\cal E$ as discussed in \cite{negyedik}. Changing this relationship
 modifies the value of the decay constant for a given radius $R$.
 The limitations for the
above mentioned relationship and for the possible radius values, and
their optimal choice ($e{\cal E} = 1.5 \kappa$ ,
$R = 0.5$ fm) were given in \cite{negyedik} comparing simulated
and experimental single-particle  distributions
 for elementary hadronic processes.

In this work we introduce the discrete final state resonances in
 the Dynamical String Model and investigate their effect on the
single-particle distributions for $2$-jet events
$e^+ e^- \to hadrons$ at $\sqrt{s} =
$ 20 - 50 GeV. Then we redetermine the string radius via the decay constant
obtained by comparing the results of numerical simulations with
experimental data for the Bose-Einstein correlation of identical pions.
Finally we present the time dependence of the pion emission.

\section*{Final state resonances}

At the very beginning of any kind of high-energy hadronic processes
 highly excited hadrons are produced. So it is plausible
 to describe these processes in terms of classical strings possessing
a continuous mass spectrum. However, the experimentally observed
particles are low energy hadrons which  definitely have a discrete
mass spectrum. Therefore the Dynamical String Model
 dealing originally only with  strings belonging to the
 continuous mass spectrum, has been improved by introducing
the discrete final state hadron resonances below the mass threshold
$1$ GeV.
These final hadron states are described by strings in the
rotating rod mode. Their properties (rest mass, decay width, the
average momentum of their decay products) are chosen according
to the phenomenology\cite{otodik}.
Thus  all the mesons
 from $\pi$ to $a_0$ are included except of those containing strange
(anti)quarks.

The treatment of decays needs a special care when the parent particles belong
to the
continuous mass spectrum and the daughters to the
discrete one. It is assumed that
the probability of choosing any species of hadron resonances is
proportional to the degree of degeneracy of this state
and the inverse of its rest mass squared
\cite{hatodik}.

The improved model described above was tested using the same
fit parameters  as for the test of its original version
in \cite{negyedik}. We concluded that the
single-particle distributions are in quantitative agreement with
the experimental data for the $2$-jet events $e^+ e^- \to hadrons$
at $\sqrt{s} \approx 20-50$ GeV\cite{hetedik}.
The effect of the discrete hadron resonances
turned out to be significant only for
 the transverse momentum
distribution  (Fig. 1).  The width of the
distribution increased due to the decay of discrete hadron resonances.

\section*{Bose-Einstein correlations}

The Bose-Einstein correlation has also been investigated in the framework
of the Dynamical String Model to extract information about the detailed
phase space structure of the simulated events $e^+ e^- \to hadrons$. We show
that the
correlation function is rather  sensitive to the value of the
decay constant. Thus it provides a better tool to fix the
parameters of the model than the single-particle distributions.

The Bose-Einstein correlation of identical bosons is the consequence
of the symmetrization of their wave function. It is reflected on the
enhancement of the number of like sign pion pairs at low momentum
differences in hadronic processes.

The correlation function is defined by
\begin{eqnarray}
\label{domi}
  C ( {\bf p}_1 , {\bf p}_2 ) =
  \frac{P ( {\bf p}_1 , {\bf p}_2 )}{P ( {\bf p}_1 ) P ( {\bf p}_2 )}
\end{eqnarray}
where $P ( {\bf p}_1 , {\bf p}_2 )$ is the two-particle momentum distribution
and $P ( {\bf p}_i )$ is the single-particle
momentum  distribution. The
product in the denominator of (\ref{domi}) can be replaced by $P_0 (
{\bf p}_1, {\bf p}_2 )$ which is the two-particle distribution in the lack
of quantum interference effects. If the source function $g ( x , {\bf p} )$
of the emitted particles is known, the correlation function in
the plane wave approximation will read\cite{nyolcadik}
\begin{eqnarray}
\label{corr}
  C ( {\bf p}_1, {\bf p}_2 ) =
  \frac{ \int d^4 x d^4 x' g ( x , {\bf p}_1 ) g ( x' , {\bf p}_2 )
  [ 1 + \cos ( p_1 - p_2 )^{\mu} ( x - x' )_{\mu} ] }
  { \int d^4 x d^4 x' g ( x , {\bf p}_1 ) g ( x' , {\bf p}_2 ) }
\end{eqnarray}
(with the Lorentz index $\mu$).
For a homogeneous source, if the momentum of the emitted
particle is independent of its space-time coordinate $x$ ,
the source function is a product
$g ( x , {\bf p} ) = \rho ( x ) g ( {\bf p} )$. Then the correlation
function is given by
\begin{eqnarray}
\label{furier}
  C ( {\bf p}_1, {\bf p}_2 ) =
  1 + | \tilde{\rho} ( {\bf p}_1 - {\bf p}_2 ) |^2
\end{eqnarray}
where the $\tilde{\rho}({\bf p})$ is the Fourier transform of
$\rho ( {\bf x} )$, and
the width of the correlation function is proportional to the inverse
of the size of the source in space.

 The source function of the produced pions
has been obtained by numerical simulation in the Dynamical String Model.
 The test for the homogeneity
of the source function shows that only the direction of the momentum
of the produced particle is independent of its space-time coordinate.
 The magnitude of the momentum
increases as the spatial distance increases  between the creation point
 of the particle
 and the  point of the initial quark-antiquark pair creation.
This can be expected from the picture of inside-outside cascade
of the hadronization in the string model, where
the low momentum states are populated
first in time and consequently in space\cite{nyolca}.

 Making use  of the simulated
source function the correlation function has been determined according to
Eq. (\ref{corr}).
Similarly to the presentation of the TPC data\cite{kilencedik}, the
correlation function was calculated as the function of the following
single variables: $q=|(p_1-p_2)^{\mu}(p_1-p_2)_{\mu}|^{1/2}$,
$q_T$ (the component of momentum difference ${\bf p}_1-{\bf p}_2$
perpendicular
to the momentum sum ${\bf p}_1+{\bf p}_2$) and $q_0$ (the energy
difference of the pion pair). In the latter two cases the restrictions
$q_0 < 0.2$ GeV and $q_T < 0.2$ GeV/c were applied, respectively.
Furthermore the same cuts were used as in the experiment
to get an appropriate sample of events, namely only the pions
with momentum $0.15$ GeV/c $< |{\bf p}| <
1.45$ GeV/c were chosen.

The calculation was carried out at first for the parameters
(taken from \cite{negyedik}) giving
the best fit to the single-particle distributions and
 practically no Bose-Einstein effect was found (Fig. 2).
The width of the fitted Gaussian function gave for the radius of the
pion source an  order of magnitude larger value than the
experimental source size in space-time. This was basicly due to
the rather large decay constant slowing down the pion emission
tremendously.

At this point one can ask whether the input parameters  can be fixed
to reproduce the experimental data for the Bose-Einstein correlation
function and the single-particle distributions simultaneously. Hence
the correlation function is more sensitive to the variation of the
decay constant than the single-particle spectra, we tried to fit
the correlation
function by an appropriate choice of  the decay constant
and then to find out the string radius and
the proper relationship between the string tension $\kappa$ and
the chromoelectric field strength ${\cal E}$.
We found
a qualitatively good description of the Bose-Einstein correlation
for $e{\cal E} = 2 \kappa$ and $R = 0.6$ fm
(Figs. 3-5). The `new' relationship and radius value
are still in the range allowed by the single-particle spectra
\cite{negyedik}. The value of the decay constant is now
$\Lambda=1.25$ fm$^{-2}$. This implies that the average 'life
time' of the expanding quark-antiquark pair in the center of
mass system is $0.9$ fm/c,
which is comparable to the 'life time' ($1.2 \pm 0.1$ fm/c) of the
string between two coloured
nucleons extracted from the rapidity distribution of protons
in the collisions of proton on proton\cite{nyolcb}.

By adjusting   the decay constant the qualitative agreement
is only achieved for the $q_0$ dependence of the correlation
function (Fig. 5). This agreement guarantees that the time
scale of the hadronization process is correct in our model.

The single-particle distributions like the transverse momentum distributions,
the Feynman $x_F$ distributions and the rapidity distributions are only
slightly modified by the new decay constant (Figs. 6-8). The calculated
average charged particle multiplicities show a softer
energy dependence than the experimental data (Fig. 9).

It is the main advantage of the Dynamical String Model that
it follows up the space-time evolution of the hadronization process
directly. So
having fixed the time scale of the process as described above,
we determined the time dependence of the pion emission in the
center of mass system of the colliding $e^+e^-$ pair (Fig. 10).
It can be seen that the pion emission rate has a maximum at about $5$ fm/c
independently of the c.m. energy.

For times $t>5$ fm/c the pion emission rate decreases exponentially
$dN/dt \propto e^{-t/T_0}$, where $T_0$ is the mean life time
of the pion source in the c.m. system of the 2-jet events (Table 1).
These values of $T_0$ cannot be compared directly to the mean
life time $\tau_0=(0.62 \pm 0.25)$ fm/c of the pion source extracted
from the dependence of the Bose-Einstein correlation function on $q_0$
and $q_T$\cite{kilencedik}, because $\tau_0$  measures the life time
of the pion
source in its rest frame. On the other hand from the ratio of the mean
life time in the laboratory and to that in the rest frame the speed of the pion
source can be estimated, $\gamma=(1-v^2/c^2)$ and $\gamma=T_0/\tau_0$.
Considering the very simplified picture
that the pion emission takes place nearby the leading (anti)quark
a kind of effective mass of the pion source can be estimated using
the speed of the source (Table 2). This estimation leads to the rest mass
$\sqrt{s}/2\gamma \approx 1.0$ GeV of the pion source.

The maximum of the pion emission rate is found at around $2$ fm/c
proper time (Fig. 11). The slope of the decay of the pion source
is determined by the characteristic proper time $2.2$ fm/c.

\section*{Conclusions}

It was established that
the Dynamical String Model improved by introducing the discrete final
state hadron resonances provides one of better quality
 single-particle spectra for the process $e^+ e^- \to hadrons$
at $\sqrt{s} \approx 20 - 50$ GeV than the original version of the
model. As the consequence of the decay of the discrete
resonances the transverse momentum
distributions are now in quantitative agreement with the
experimental data for small transverse momenta  up to $p_T < 2$ GeV/c.

The simulations in the framework of the Dynamical String Model
 have shown that the
Bose-Einstein correlation for the
  input parameters representing the best fit
to the single-particle distributions (obtained in \cite{negyedik})
 is in contradiction
with the TPC experiment. It has also been  established that
the simulated Bose-Einstein
correlation is rather sensitive to the value of the decay constant.
Making use of the ambiguities
 of the hadronic string - flux tube analogy, it was possible to choose
the string radius $R = 0.6$ fm and achieve a decay constant
providing a rather good agreement of both the simulated Bose-Einstein
correlation and the single particle distributions with the
experimental data. The appropriate string decay constant implies
the life time $0.9$ fm/c of the hadronic string which is close to
the value obtained in Ref.\cite{nyolcb}.

Working with this decay constant that allows to recover the observed
$q_0$ dependence of the Bose-Einstein correlation, the time scale
of the hadronization process was fixed in a reliable way. Then the
pion emission rate and the mean life time of the pion source were
determined. It was established that the pion source builds up
nearly in the first $5$ fm/c in the laboratory frame ($2$ fm/c in
proper time), independently
of the c.m. energy. Its mean life time is linearly rising with the
c.m. energy. Comparing the mean life time detemined by the pion
emission rate to the mean life time extracted from the
Bose-Einstein correlation an effective rest mass ($1.0$ GeV) of the
pion source nearby the leading (anti)quark can be found.

\section*{Acknowledgement}

One of the authors (B.I.) wishes to express his thanks to J. Salo and F. Kun
for helping in the data analysis and B. M\"uller for drawing his
attention to Ref.\cite{nyolcb}.
This work was supported by the EC project ERB-CIPA-CT92-4023,
proposal 3293 and by the Hungarian Research Fund OTKA 2192/91.

\begin{table}

\begin{tabular}{|c|c|c|c|}  \hline
 c.m. energy (GeV) & $22$ & $35$ & $50$  \\ \hline
 $T_0$ (fm/c) & $7.35 \pm 0.13$ & $10.1 \pm 0.1$ & $13.2 \pm 0.15$  \\ \hline
\end{tabular}

\caption{
$T_0$ parameters of the exponential decay of the pion emission
rate (see text) for different c.m. energies.
}
\end{table}

\begin{table}

\begin{tabular}{|c|c|c|c|}  \hline
 c.m. energy (GeV) & $22$ & $35$ & $50$  \\ \hline
 m (GeV) & $0.93 \pm 0.02$ & $1.07 \pm 0.01$ & $1.17 \pm 0.01$  \\ \hline
\end{tabular}

\caption{
The "effective mass" of the pion source
(see text) for different c.m. energies.
}
\end{table}

\begin{figure}
\epsfxsize=1.0\columnwidth
\epsffile{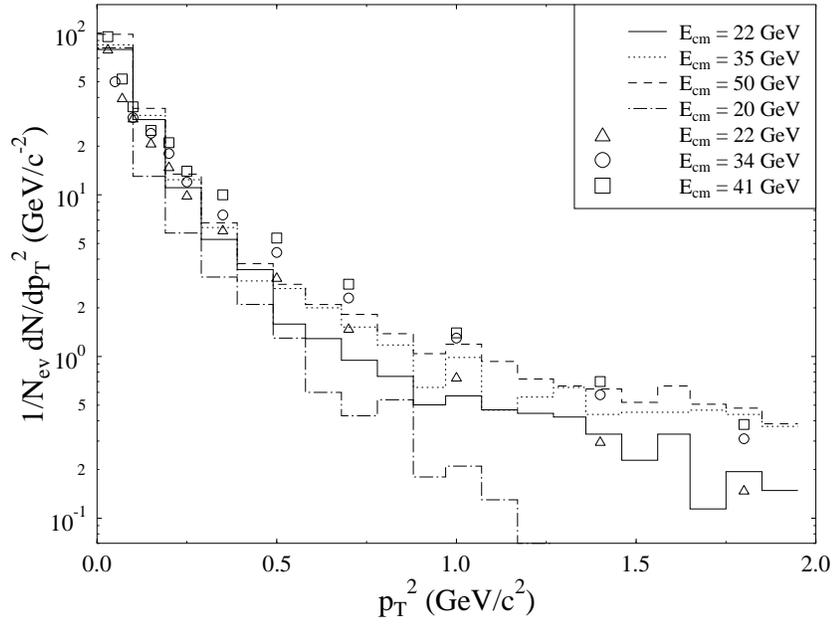}
\caption{
Transverse momentum distributions.
Lines (full, dotted, dashed)
represent the simulations of the model with discrete resonances and using
the original input parameters. Dotted-dashed line indicates the data
obtained by the model without discrete resonances[1].
Markers are the experimental
data taken from Ref. [12].
}
\end{figure}

\begin{figure}
\epsfxsize=1.0\columnwidth
\epsffile{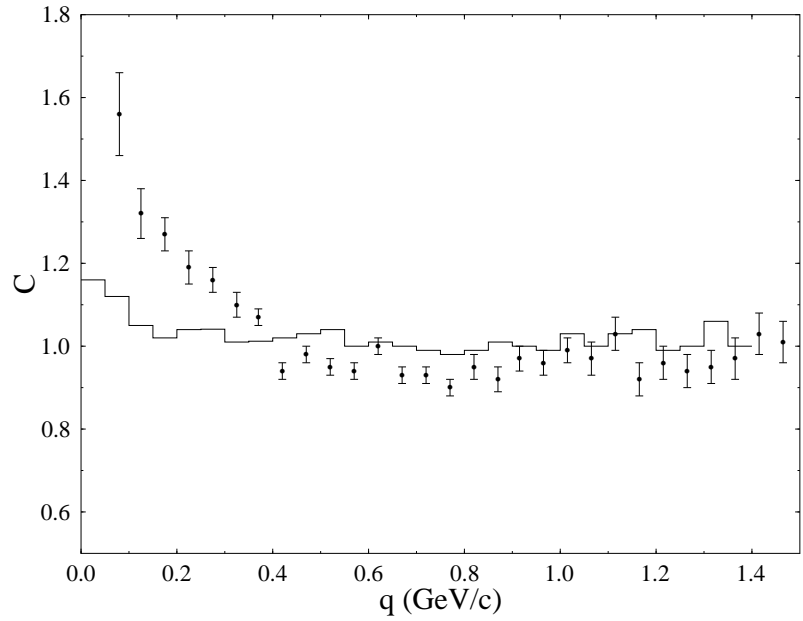}
\caption{
Correlation function $C$ of like sign pion pairs as a function of the
four momentum difference ($q=|p_1-p_2|$). The simulated data of the
model with the original input parameters are represented by the full
line. Markers are for the experimental data at $29$ GeV of the TPC
collaboration[11].
}
\end{figure}

\begin{figure}
\epsfxsize=1.0\columnwidth
\epsffile{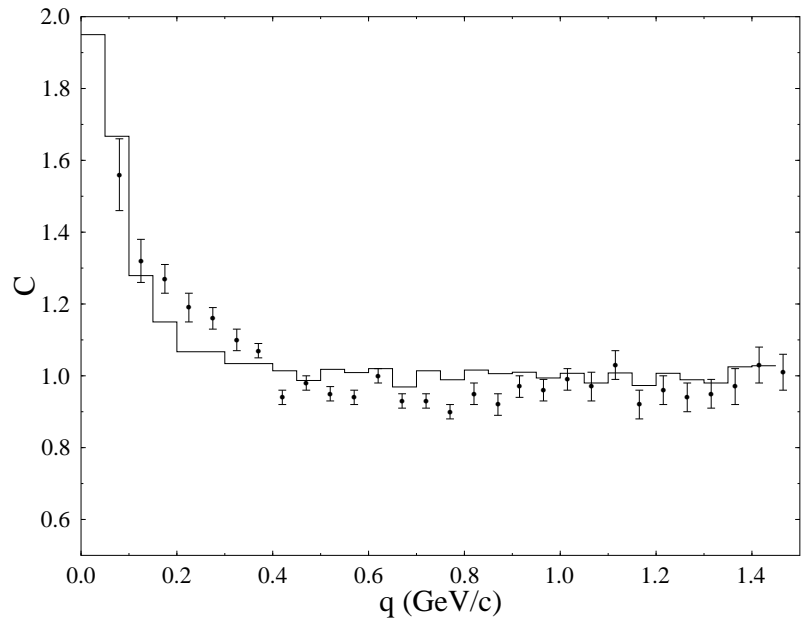}
\caption{
Correlation function $C$ of like sign pion pairs as a function of the
four momentum difference ($q=|p_1-p_2|$). The simulated data of the
model with the input parameters giving the best fit are represented by the full
line. Markers are for the experimental data of the TPC collaboration[11].
}
\end{figure}

\begin{figure}
\epsfxsize=1.0\columnwidth
\epsffile{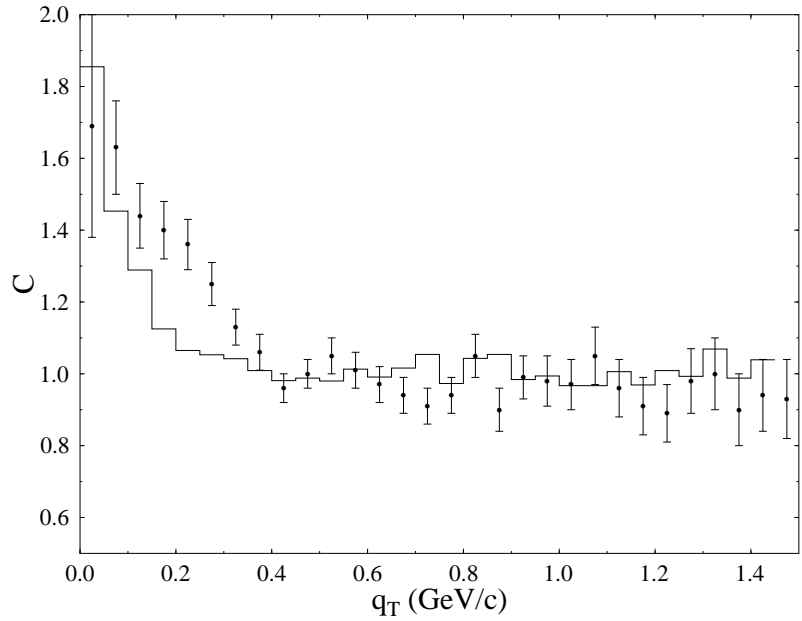}
\caption{
Correlation function $C$ of like sign pion pairs as a function of
$q_T$, which is the component of the three momentum difference
${\bf p}_1 - {\bf p}_2$ perpendicular to the momentum sum
${\bf p}_1 + {\bf p}_2$, for the energy difference $q_0 < 0.2$ GeV/c
(see Fig. 3. for notation).
}
\end{figure}

\begin{figure}
\epsfxsize=1.0\columnwidth
\epsffile{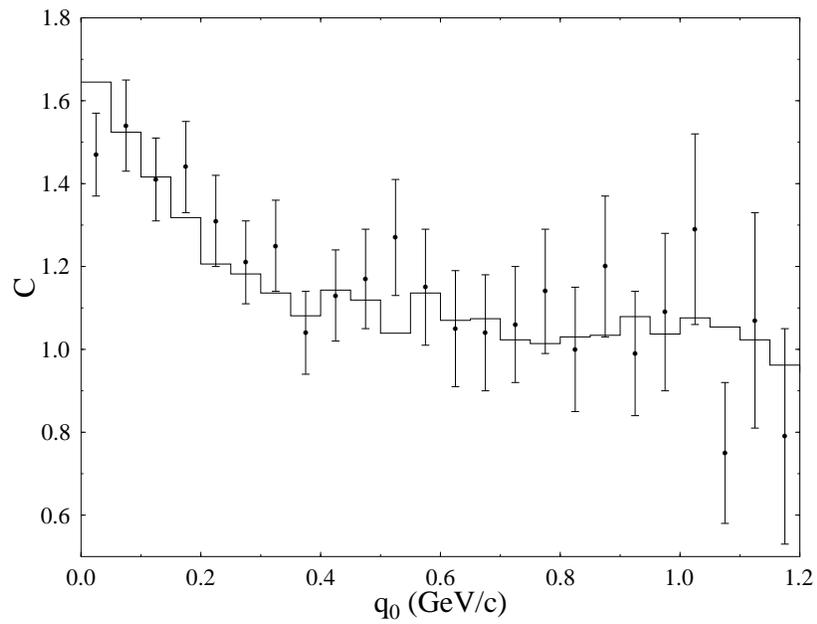}
\caption{
Correlation function $C$ of like sign pion pairs as a function of the
energy difference $q_0$
, for $q_T < 0.2$ GeV/c
(see Fig. 3. for notation).
}
\end{figure}

\begin{figure}
\epsfxsize=1.0\columnwidth
\epsffile{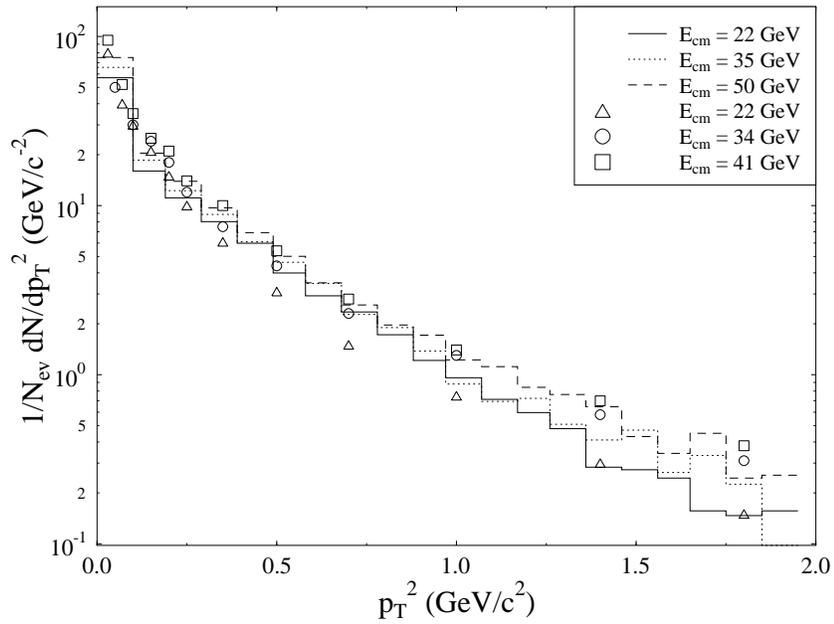}
\caption{
Transverse momentum distributions.
Lines
represent the results of simulations with discrete resonances and using
the input parameters fitted to the correlation functions.
Markers are for the experimental
data taken from Ref. [12].
}
\end{figure}

\begin{figure}
\epsfxsize=1.0\columnwidth
\epsffile{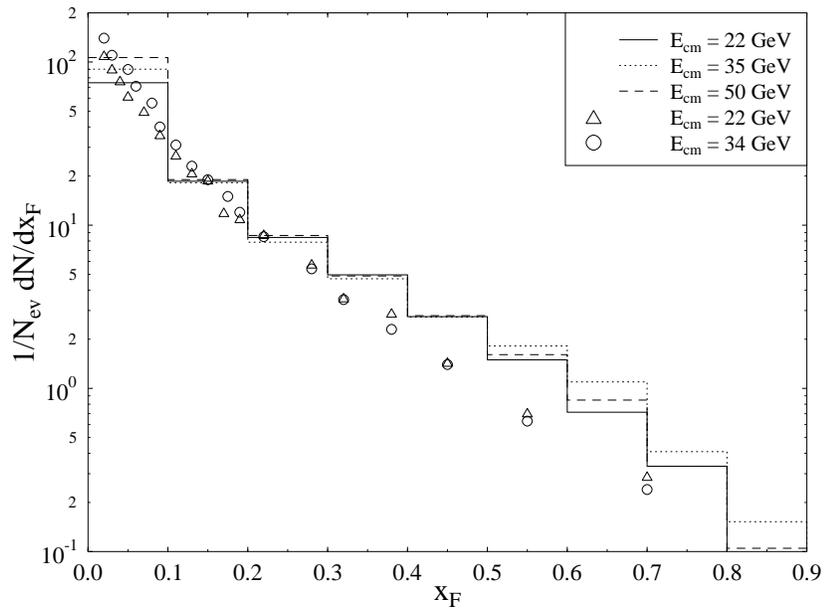}
\caption{
The Feynmann $x_F$ ($x_F = \frac{2 p_L}{E_{cm}}$) distributions.
(Lines and markers are the same as in Fig. 6.)
}
\end{figure}

\begin{figure}
\epsfxsize=1.0\columnwidth
\epsffile{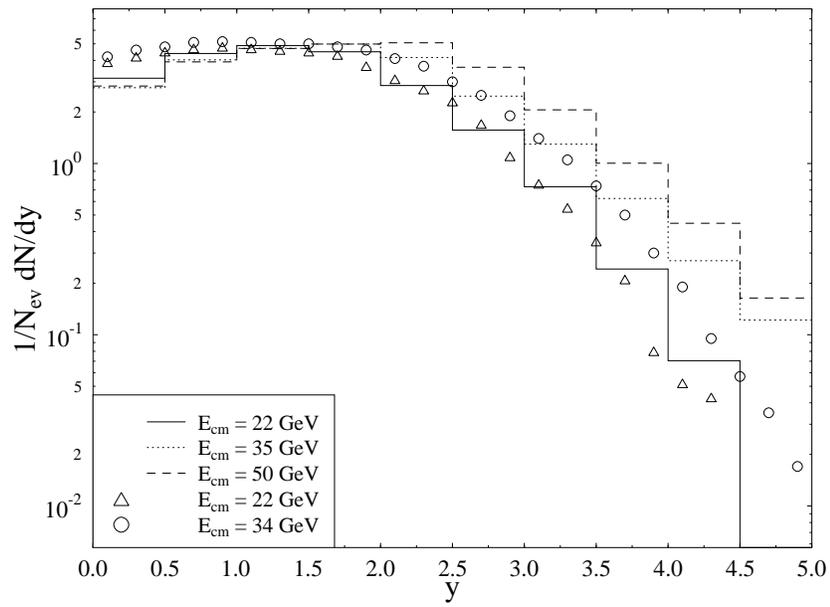}
\caption{
Rapidity distributions.
(Lines and markers are the same as in Fig. 6.)
}
\end{figure}

\begin{figure}
\epsfxsize=1.0\columnwidth
\epsffile{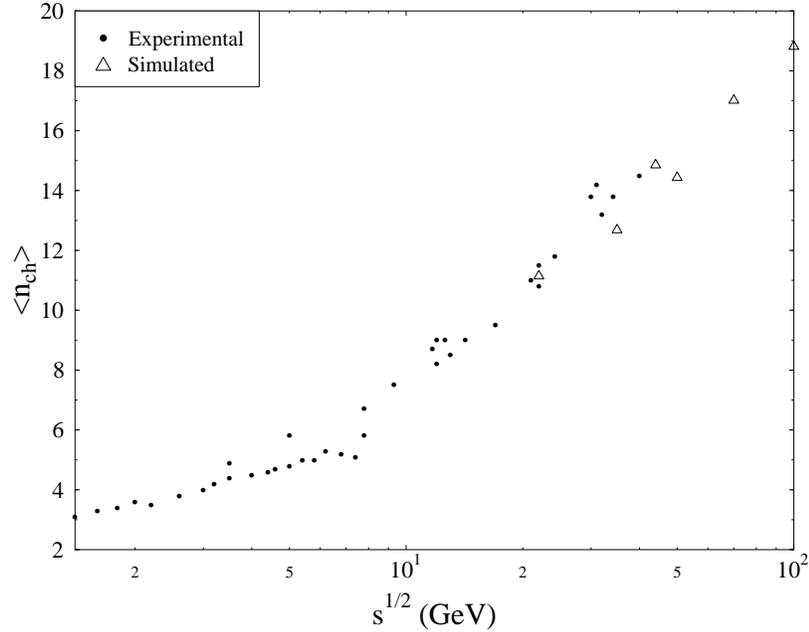}
\caption{
The average multiplicity as the function of the c.m. energy.
}
\end{figure}

\begin{figure}
\epsfxsize=1.0\columnwidth
\epsffile{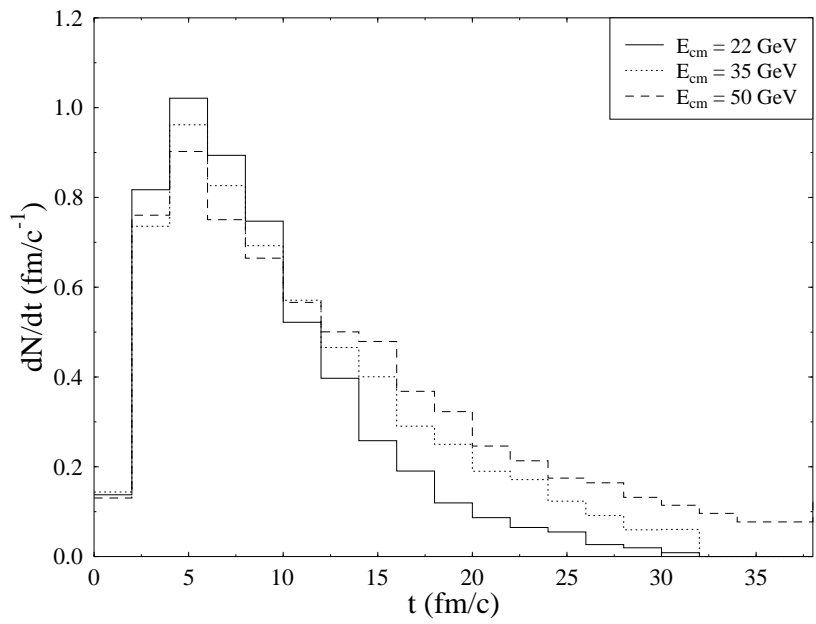}
\caption{
The time distribution of the pion emission in the c.m. system of the
two-jet events (Lines are the same as in Fig. 6.)
}
\end{figure}

\begin{figure}
\epsfxsize=1.0\columnwidth
\epsffile{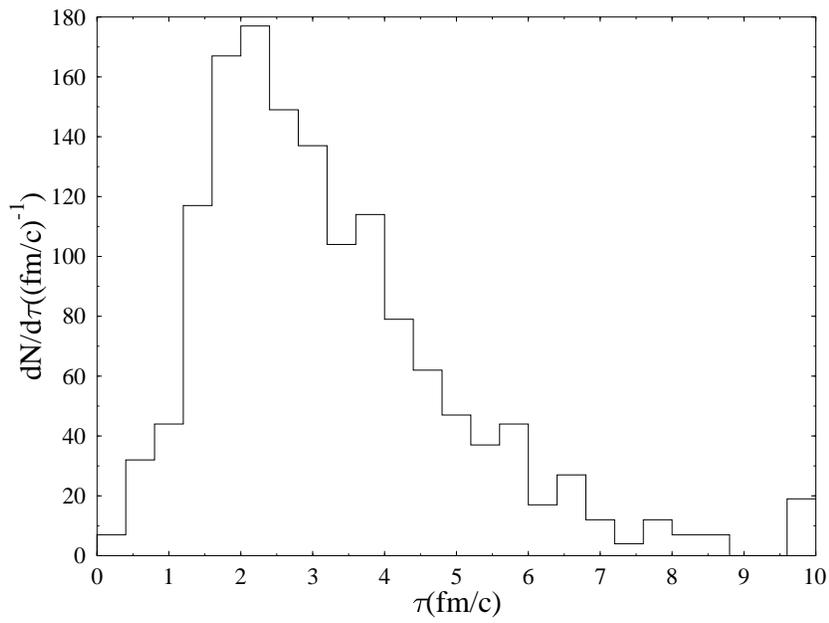}
\caption{
The proper time distribution of the pion emission at c.m. energy $29$
GeV.
}
\end{figure}

\end{document}